\documentclass[aip,sd,amsmath,amssymb,preprint,preprint]{revtex4-1}
\usepackage{graphics}
\usepackage{times}
\pdfoutput=1 
\begin{document}

\title{Effective-medium-cladded~dielectric~waveguides for terahertz communications}
\author{Weijie~Gao}
\
\affiliation{Terahertz Engineering Laboratory, School of Electrical and Electronic Engineering, The University of Adelaide, South Australia 5005, Australia}

\author{Xiongbin~Yu}
\affiliation{Graduate School of Engineering Science, Osaka University, 1-3 Machikaneyama, Toyonaka, Osaka 560-8531, Japan}

\author{Masayuki~Fujita}
\email{fujita@ee.es.osaka-u.ac.jp}
\affiliation{Graduate School of Engineering Science, Osaka University, 1-3 Machikaneyama, Toyonaka, Osaka 560-8531, Japan}

\author{Tadao~Nagatsuma}
\affiliation{Graduate School of Engineering Science, Osaka University, 1-3 Machikaneyama, Toyonaka, Osaka 560-8531, Japan}

\author{Christophe~Fumeaux}
\affiliation{Terahertz Engineering Laboratory, School of Electrical and Electronic Engineering, The University of Adelaide, South Australia 5005, Australia}

\author{Withawat~Withayachumnankul}
\email{withawat@adelaide.edu.au}
\affiliation{Terahertz Engineering Laboratory, School of Electrical and Electronic Engineering, The University of Adelaide, South Australia 5005, Australia}
\date{\today}


\begin{abstract}
\textbf{Terahertz communications is a promising modality for future short-range point-to-point wireless data transmission at rates up to terabit per second. A milestone towards this goal is the development of an integrated transmitter and receiver platforms with high efficiency. One key enabling component is a planar waveguiding structure with wide bandwidth and low dispersion. This work proposes substrate-less all-dielectric waveguides cladded by an effective medium for low-loss and low-dispersion terahertz transmission in broadband. This self-supporting structure is built solely into a single silicon wafer with air perforations to mitigate significant absorptions in metals and dielectrics at terahertz frequencies. The realized waveguides can cover the entire 260--400~GHz with single dominant modes in both orthogonal polarizations. The simulation shows that for the $E_{11}^{x}$ mode the attenuation ranges from 0.003 to 0.024 dB/cm over the entire band, while it varies from 0.008 to 0.023 dB/cm for the $E_{11}^{y}$ mode. Limited by the measurement setup, the maximum error-free data rate of 28 Gbit/s is experimentally achieved at 335 GHz on a 3-cm waveguide. We further demonstrate the transmission of uncompressed 4K-resolution video across this waveguide. This waveguide platform promises integration of diverse active and passive components. Thus, we can foresee it as a potential candidate for the future terahertz integrated circuits, in analogy to photonic integrated circuits at optical frequencies. }
\end{abstract}


\maketitle

Data traffic has grown exponentially with a dramatically increasing proportion of the world population accessing online services.~\cite{cisco} This trend has been exacerbated by the rapid growth in usage of high-quality multimedia content with huge data volume. Therefore, higher channel capacities in wireless transmission are in demand to support not only mobile links but also links among static nodes in the backbones. However, an increased data throughput would pressure the already congested microwave and millimetre-wave channels.~\cite{Nagatsuma2016a} According to the Shannon-Hartley theorem, the channel capacity can be improved by increasing the bandwidth of the carrier, subject to signal quality.~\cite{goldsmith2005wireless} 

Bridging electronics and optics, the terahertz spectrum that spans the frequency range from 0.1 to 10~THz is an ideal candidate to support high-speed wireless line-of-sight transmission owing to its ultra-wide absolute bandwidth.~\cite{Nagatsuma2016a} In theory, the prospective data rate carried by a terahertz channel can reach 1~Tbit/s over the distance of several kilometres.~\cite{koenig2013wireless} However, reaching this theoretical capacity calls for solutions to several challenges, including the significant power losses in integrated transmitter and receiver components that compromise signal quality. This issue can be mitigated by using bulky discrete components with trade-offs in dispersion, fabrication complexity, and compactness.~\cite{gallot2000terahertz,mcgowan1999propagation} Hence, to realize integrated terahertz communications systems with enhanced performance, planar interconnects with low loss and low dispersion in broadband are vital.~\cite{Withayachumnankul2018a,Yan2013,wu2006a,Taeb2016,BCB2018,Zhu2016a,Fujita2015b,Fujita2016,Yu2019}

At~terahertz~frequencies,~conventional~planar~metallic~transmission~lines,~such~as~microstrip lines and coplanar~waveguides,~impose~significant~transmission~losses~in~the~order~of~several~\mbox{dB/cm}.~\cite{Frankel1991,Zhang2005,murano2017low} This is caused by their strong field confinement, together with substantial ohmic and dielectric losses at high frequencies. The same issues apply to the substrate integrated waveguides that have been widely used in millimetre-wave applications.~\cite{Wu2003} An alternative is substrate-integrated image guide (SIIG) technology that employs metal-grounded dielectric waveguides.~\cite{wu2006a,Taeb2016} This type of waveguide could achieve an average transmission loss of 0.45~dB/cm from 85 to 90~GHz,~\cite{wu2006a} and 0.35~dB/cm from 110 to 170~GHz.~\cite{Taeb2016} While these metal-based guiding structures are efficient at microwave and millimetre-wave frequencies, they are not suitable to terahertz integrated systems due to the increased losses and bandwidth restrictions. 

Inspired by guided-wave optics, all-dielectric waveguides with low transmission losses have been investigated at terahertz frequencies in the past. For example, dielectric waveguides were proposed with an average propagation loss of 0.087~dB/cm over two bands (90--140 GHz and 140--220~GHz).~\cite{Ali2014} In spite of their low losses and low dispersion, such bare dielectric waveguides have limitations in realizing integrated systems since they lack self-support and support for other components.~To improve the integrability, silicon-on-insulator (SOI) waveguides were proposed,~\cite{BCB2018,Zhu2016a} where the average loss was 0.46~dB/cm over 500 to 580~GHz.~\cite{BCB2018} With structural resemblance, a dielectric microstrip line concentrates the energy in the low-index oxide spacer.~\cite{Zhu2016a} In this case, the loss varied from 0.23 to 1.2~dB/cm over 750 to 925~GHz.~\cite{Zhu2016a} However, the losses imposed by additional supporting materials are not negligible.~\cite{lee2001fabrication} To achieve an integrated waveguide platform with low loss and structural simplicity, all-dielectric 2D photonic crystal waveguides built from a single silicon wafer were proposed.~\cite{Fujita2015b,Yu2019} Silicon, as the only constituent material, has extremely low dissipation. As such, a propagation loss of less than 0.1~dB/cm could be demonstrated from 319 to 337~GHz,~\cite{Fujita2015b} while another similar design could yield an enhanced bandwidth of 324--361~GHz with comparable losses.~\cite{Yu2019} However, these photonic crystal waveguides have relatively narrow bandwidths and strong in-band dispersion related to the intrinsic photonic bandgap (PBG) phenomenon.
  
In this paper, we propose a class of self-supporting all-dielectric waveguides. The proposed substrate-less waveguides are constructed solely from a high-resistivity silicon wafer that has exceptionally low loss and low dispersion at terahertz frequencies.~\cite{Dai2004} A silicon waveguide core is surrounded by in-plane effective-medium claddings, which are realized by periodic perforation of the silicon slab. In a stark contrast to photonic crystal waveguides, the perforation period of the proposed waveguide is in the deep subwavelength region. In this way, the perforated medium behaves like a homogeneous material rather than a bandgap material.~\cite{Cheben2018} Compared to the SIIG waveguides, the proposed design is free from metal thus achieving lower transmission losses, while able to sustain two orthogonal dominant modes. Compared to the SOI waveguides, the proposed design is self-supporting and does not require additional materials that would contribute to propagation losses.
\begin{figure}[!htb]
	\centering
	\includegraphics{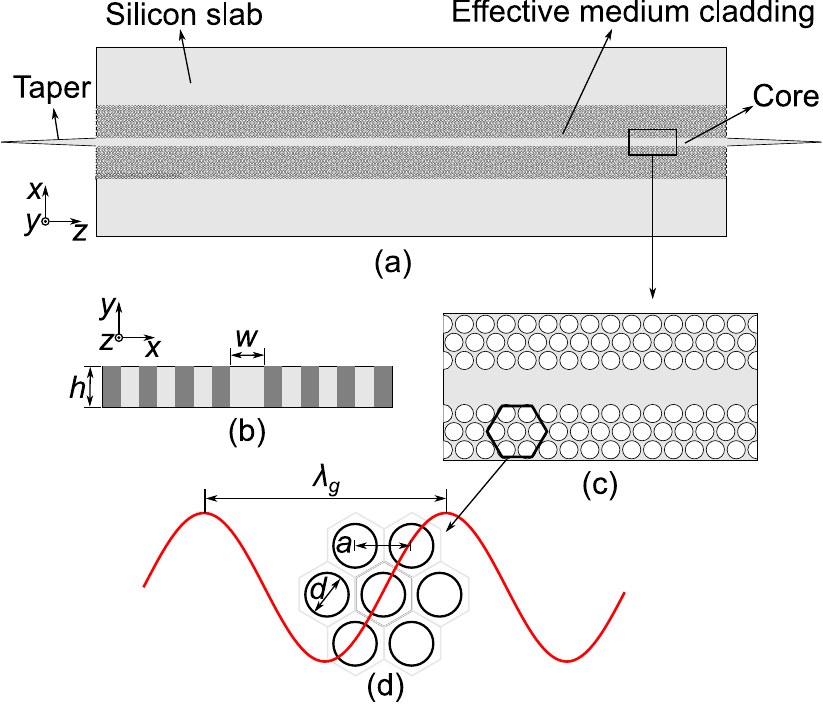}
	\centering\caption{\textbf{Effective-medium-cladded dielectric waveguide.} (a) Top view, (b) magnified cross-sectional view, (c) magnified view of waveguide core and claddings, and (d) hexagonal lattice of the effective medium cladding with perforation period $a$ and hole diameter $d$. The period $a$ is a quarter of the shortest guided wavelength $\lambda_g$.~This is in contrast to photonic crystal cladding that requires a periodicity close to half a guided wavelength. The tapers are coupling structures to be inserted into the hollow metallic waveguides required for simulations and measurements. The dimensions of the final design are: $h$ = 200~$\mu$m, $w$~=~160~$\mu$m, $a$ = 100~$\mu$m, $d$ = 90~$\mu$m. It is noted that the lateral unperforated silicon regions are for handling purpose, and they do not interfere with the guided modes. }
	\label{fig:SWG}
\end{figure}

As shown in Fig.~\ref{fig:SWG}(a--c), the proposed integrated all-dielectric waveguide comprises a waveguide solid core and in-plane effective medium claddings, both built monolithically into a silicon slab. The unperforated areas are for handling purposes and do not interfere with the propagation modes of the waveguide. The silicon slab with a thickness of 200~$\mu$m has a relative permittivity $\epsilon_{\rm Si}$ of 11.68 (an equivalent refractive index $n_{\rm Si}$ of 3.418) and a resistivity of $20$~k$\Omega$-cm.~Such high-resistivity silicon is nondispersive and offers extremely low loss, with power absorption coefficients less than 0.01/cm in the entire terahertz band.~\cite{Dai2004} The effective medium is realized by periodically perforating the silicon slab with cylindrical air holes in a hexagonal lattice, with a period $a$ much smaller than the shortest guided wavelength $\lambda_g$ as shown in Fig.~\ref{fig:SWG}(d). The effective relative permittivity of the effective medium falls between those of air and silicon, and is determined by the perforation period $a$ and the hole diameter $d$. Owing to the refractive index contrast between the core and the claddings, i.e., between silicon and effective medium in plane, and silicon and air out of plane, the waves can be confined within the waveguide core by total internal reflections. Due to the low-index claddings in both transverse dimensions, the waveguide supports two orthogonal fundamental modes $E_{11}^{x}$ and $E_{11}^{y}$, namely with the polarization for the $E_{11}^{x}$ mode being parallel to the slab, while being perpendicular to the slab for the $E_{11}^{y}$ mode.~Given properly designed waveguide width, perforation period, and hole diameter, the waveguide can operate with a single mode over 260 to 400 GHz (WR-2.8 band) for each polarization.

Despite their similar appearances, the proposed effective-medium-cladded dielectric waveguides are fundamentally different from the photonic crystal waveguides~\cite{Fujita2015b,Yu2019} in terms of operation mechanism.~\cite{Cheben2018,Mosallaei2002} Specifically, the guidance of the photonic crystal waveguides relies on the bandgap effects. The bandgap material prohibits propagation modes by means of interference and thus confines the waves within the core. The realization of the bandgap materials is based on a variation in refractive indices of two alternating dielectric materials with the period equal to or longer than half of the guided wavelength to cause destructive interference. In contrast, the effective-medium-cladded waveguides rely merely on the total internal reflection attributed to the refractive index contrast between the waveguide core and the claddings. In this case, the purpose of the deeply sub-wavelength perforation is to lower the effective refractive index of the silicon \mbox{slab} to create an effective medium.~\cite{Cheben2018,Birman2009} The considerations for the effective medium and waveguide modal analysis will be discussed in the following.      

As shown in Fig.~\ref{fig:SWG}(d), the effective medium claddings are in the form of a hexagonal array of cylindrical thru-holes perforated into the silicon. The relative permittivity of the effective medium can be obtained by Maxwell--Garnett approximations,~\cite{Subashiev2006,Birman2009} which are polarization-dependent. These approximations are valid only when the periodicity is subwavelength, where wave diffraction and scattering are negligible. In this way, the effective medium can be treated as a homogeneous material with an equivalent anisotropic permittivity tensor with respect to the macroscopic electromagnetic field.~\cite{Cheben2018,Birman2009} Specifically, the relative permittivities for the $E$-field perpendicular and parallel to the axis of the cylindrical air hole are defined as $\epsilon_{x}$ and $\epsilon_{y}$ respectively. They are given as~\cite{Subashiev2006}
\begin{eqnarray}
\epsilon_{x}&=&\epsilon_{\rm Si}\frac{(\epsilon_{0}+\epsilon_{\rm Si})+(\epsilon_{0}-\epsilon_{\rm Si})\zeta}{(\epsilon_{0}+\epsilon_{\rm Si})-(\epsilon_{0}-\epsilon_{\rm Si})\zeta}\;, 
\label{eq1}\\
\epsilon_{y}&=&\epsilon_{\rm Si}+(\epsilon_{0}-\epsilon_{\rm Si})\zeta\;. 
\label{eq2}
\end{eqnarray}
where $\epsilon_{0}$ and $\epsilon_{\rm Si}$ are the relative permittivities of the air and silicon respectively, while $\zeta$ represents the fill factor of the air volume in silicon. For a hexagonal lattice, the fill factor can be calculated from ($\pi$$d^2$)$/(2\sqrt{3}a^2$), where $d$ is the diameter of the cylindrical air hole and $a$ is the period of the lattice. According to Eqs.~\ref{eq1} and \ref{eq2}, the effective relative permittivities for both polarizations are monotonously decreasing with the fill factor and varying between the values of silicon and air.

In addition to the subwavelength periodicity, the design process for the effective-medium claddings needs to account for the following considerations: (i) single mode operation, (ii) wave confinement and (iii) structural integrity. Since the cutoff frequencies of the higher-order modes are affected by the claddings, the effective relative permittivity must be selected properly to enable a single-mode operation for each polarization over the desired band. Specifically, a lower relative permittivity of the claddings would generate a lower propagation constant thus pushing the cutoff frequencies of higher-order modes upward. To obtain a well guided wave with its power mostly concentrated within the core, a higher refractive index contrast between the core and the claddings is preferred.~\cite{Marcatili1969,Okamoto2007} These two conditions can be satisfied with an effective medium having a larger portion of air. As a trade-off, a larger air portion in the claddings compromises mechanical strength of the platform.

By considering all those factors mentioned above, the periodicity $a$ is selected as 100~$\mu$m in this design, corresponding to about 0.46$\lambda_{\rm Si}$, where $\lambda_{\rm Si}$ is the wavelength in bulk silicon at 400~GHz. This estimation is the worst case, since the guided wavelength is always larger than that in the bulk material. The hole diameter is chosen as 90~$\mu$m generating $\epsilon_{x}$ and $\epsilon_{y}$ of 2.75 and 3.84 respectively. These effective permittivities are much lower compared with the relative permittivity of bulk silicon of 11.68. The hexagonal hole lattice is adopted in order to reinforce the mechanical strength of the platform.
  
\begin{figure}[!h]
	\centering
	\includegraphics{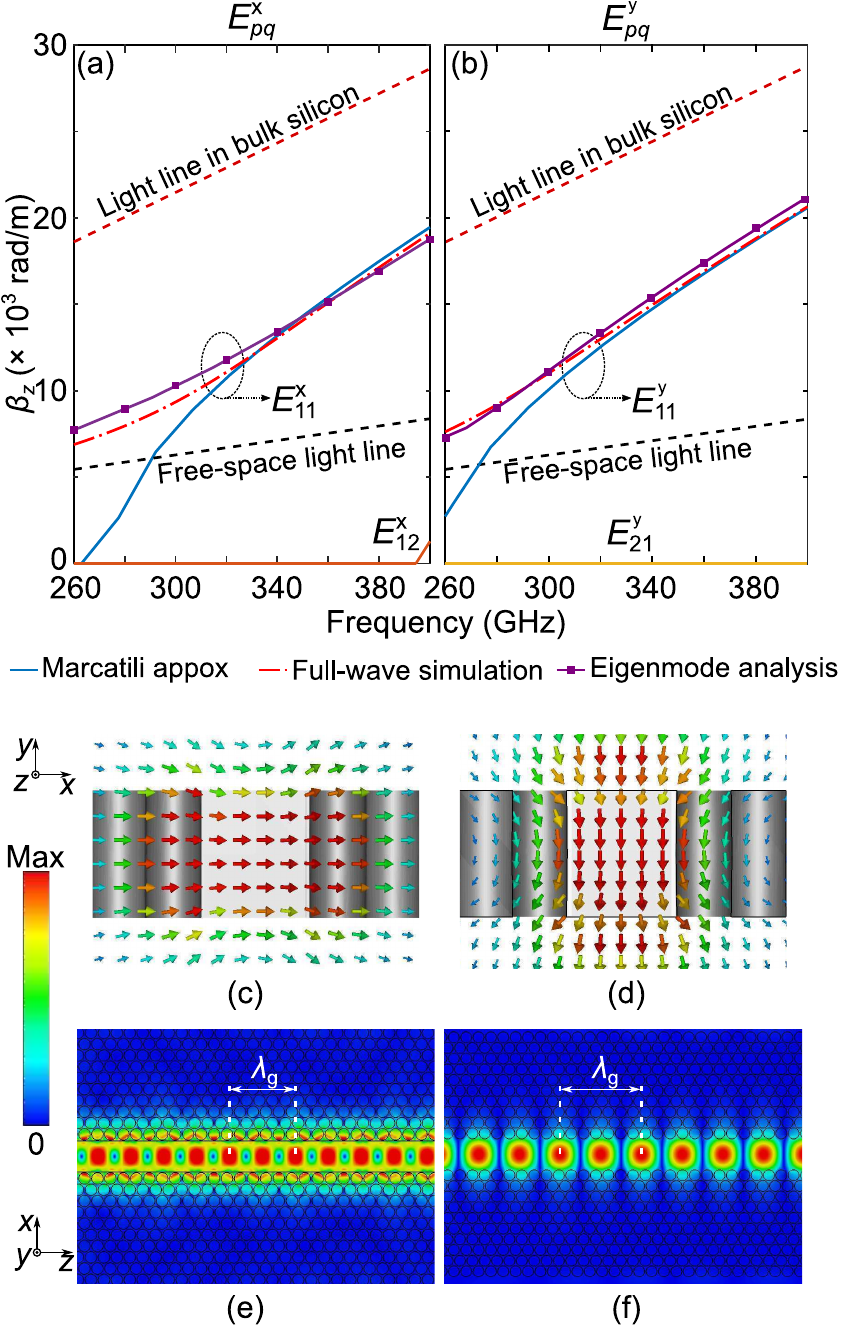}
	\caption{\textbf{Characteristics of the waveguide.} Dispersion characteristics for (a) $E_{pq}^{x}$ and (b) $E_{pq}^{y}$ modes. Simulated $E$-field distributions in linear scale at 330 GHz for (c, e) $E_{11}^{x}$ and (d, f) $E_{11}^{y}$ modes in the cross-sectional and top views, respectively. The propagation constants in the case of full-wave simulation and Eigenmode analysis are obtained from CST Microwave Studio. The propagation constants for higher-order modes are obtained from Marcatili approximation, and these modes remain below the cutoff in this frequency range.}
	\label{fig:dispersion}
\end{figure}
\begin{figure}[!h]
	\centering
	\includegraphics{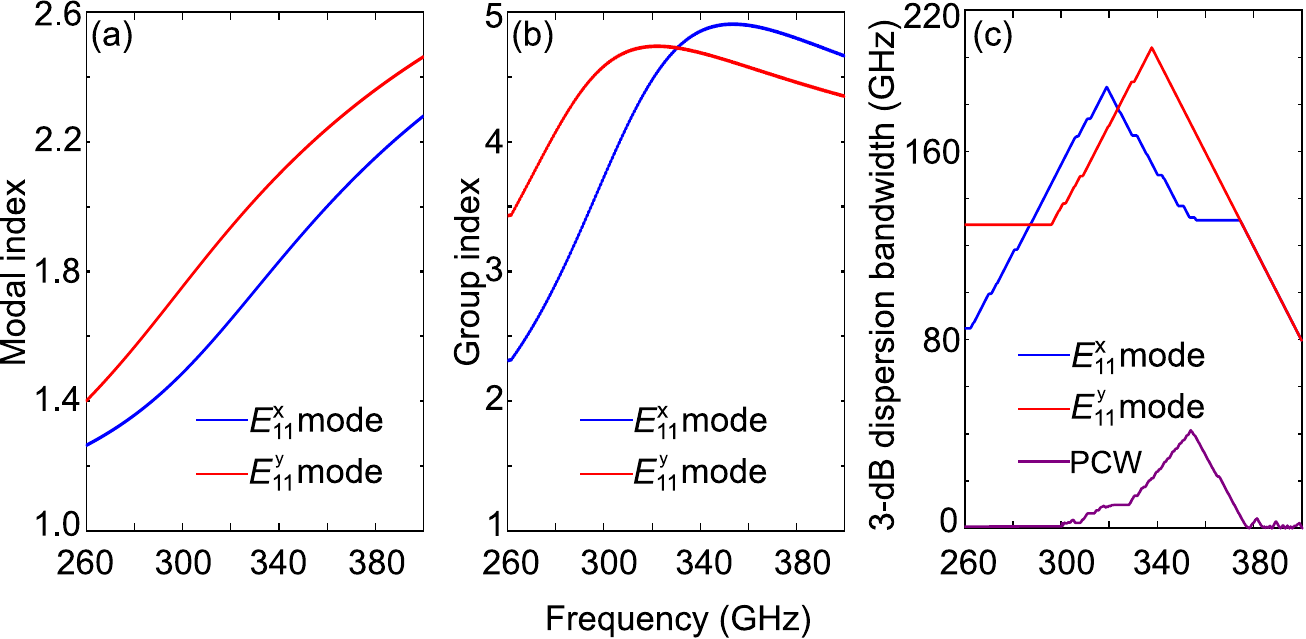}
	\caption{\textbf{Simulated dispersion characteristics.} (a) Modal indices, (b) group indices, and (c) calculated 3-dB dispersion bandwidth for the effective-medium-cladded waveguide and photonic crystal waveguide. The lengths of the effective-medium-cladded waveguide and the bandwidth-enhanced photonic crystal waveguide~\cite{Yu2019} for 3-dB bandwidth calculation are 3~cm.}
	\label{fig:effr}
\end{figure}
Marcatili's theory~\cite{Marcatili1969} is applied here to investigate the dispersion characteristics of the waveguide. According to this theory,~\cite{Marcatili1969} the propagation constant $\beta_z$ is a function of the operation wavelength $\lambda$, the width $w$ and the height $h$ of the waveguide core, as well as the relative permittivities of the core and effective medium claddings. In the specific design presented in this paper, the height of the waveguide core is 200~$\mu$m, while its width is selected as 160~$\mu$m. Given the effective relative permittivity tensor of the claddings, the dispersion characteristics can be theoretically obtained. It is noted that Marcatili's theory has limitations in accurately predicting the dispersion behaviour of the fundamental modes at low frequencies owing to the mode approximations.~\cite{Marcatili1969} Therefore, Eigenmode analysis and full-wave simulations are performed in CST Microwave Studios to complement the analytical results. 

As shown in Fig.~\ref{fig:dispersion}(a) and (b), the results from the three approaches for the fundamental modes are in good agreement, and converge at high frequencies. Within the operation frequency range from 260 to 400~GHz, the waveguide can work with low dispersion in single dominant modes for both polarizations. In addition, we can infer that the waves are tightly confined, as all the propagation constants for the $E_{11}^{x}$  and $E_{11}^{y}$ modes are well above the free-space light line. In addition, the higher-order mode appearing above 395 GHz for the horizontal polarization will not interfere with the dominant modes, since they have propagation constants below the light line, i.e., the wave confinement is extremely weak. However, the higher-order mode might cause slight radiation loss and reduce the waveguide efficiency. In this way, the bandwidth of the waveguide might need to be reduced to 260--395 GHz in some particular scenarios. The simulated $E$-field distributions for both dominant modes at 330 GHz, i.e., at the centre frequency, are shown in Fig.~\ref{fig:dispersion}(c--f). It is observed that most of the power is concentrated within the waveguide core with evanescent fields extending a few rows into the claddings. Based on the dispersion characteristics, the effective modal indices for the two dominant modes can be obtained. As shown in Fig.~\ref{fig:effr}(a), the effective modal index for the $E_{11}^{x}$ mode is lower than that for the $E_{11}^{y}$ mode. This disparity results from the differences between the width and height of the waveguide core and between the in-plane and out-of-plane claddings.~\cite{chrostowski2015silicon} Based on these modal indices, the periodicity of the effective medium amounts to only  0.3$\lambda_g$, where $\lambda_g$ is the shortest guided wavelength at 400~GHz. Derived from the modal indices, the group index for the $E_{11}^{x}$ mode is found to be higher than that for the $E_{11}^{y}$ over 260--335 GHz, while the trend inverses over 335--400 GHz as shown in Fig.~\ref{fig:effr}(b). 

The calculated 3-dB dispersion bandwidths based on the group delay~\cite{fiber} of a 3 cm-waveguide for both the $E_{11}^{x}$ and $E_{11}^{y}$ modes are shown in Fig.~\ref{fig:effr}(c). Owing to the low dispersion, the 3-dB bandwidth for each carrier frequency is high, i.e., the waveguide can support high-speed communications with ultra-broad bandwidth. The 3-dB dispersion bandwidth ranges from 80 to 185 GHz for the $E_{11}^{x}$ mode, while it varies from 80 to 190 GHz for the $E_{11}^{y}$ mode over the entire frequency range. The discrepancies of the bandwidth for the two modes are due to a slight difference in the dispersion characteristics. Compared to the bandwidth-enhanced photonic crystal waveguide (PCW),~\cite{Yu2019} the calculated 3-dB dispersion bandwidths of the proposed waveguide are much higher. This performance improvement is attributed to its extremely low in-band dispersion.
\begin{figure}[!t]
	\centering
	\includegraphics{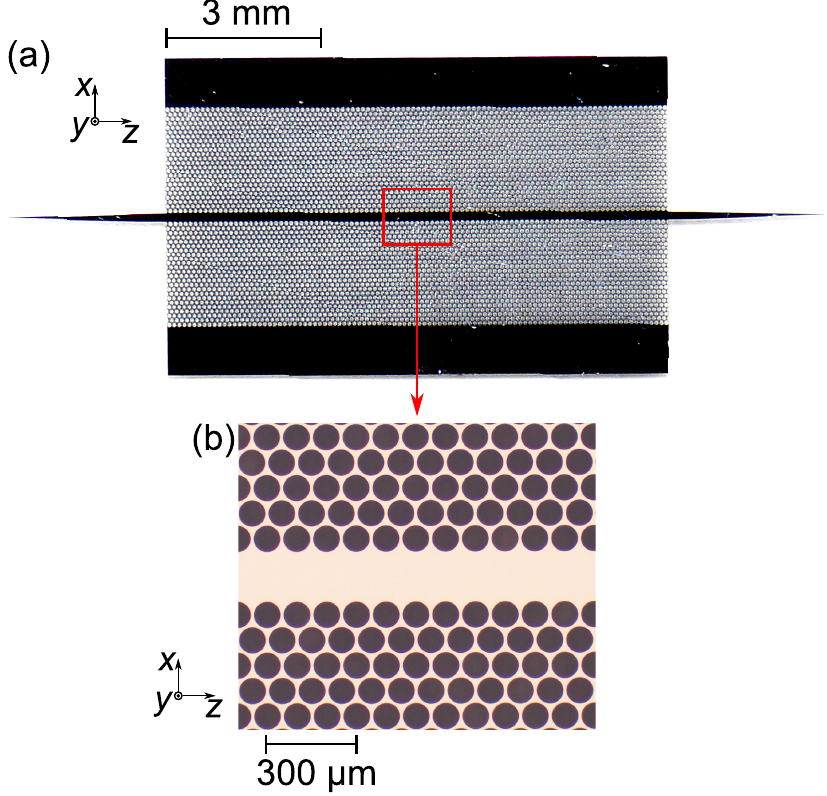}
	\caption{\textbf{Fabricated effective-medium-cladded dielectric waveguide.} (a) 1 cm waveguide sample, and (b) magnified view of the effective-medium claddings. The design parameters are given in the caption of Fig.~\ref{fig:SWG}.}
	\label{fig:fabrication}
\end{figure} 

\begin{figure}[!t]
	\centering
	\includegraphics{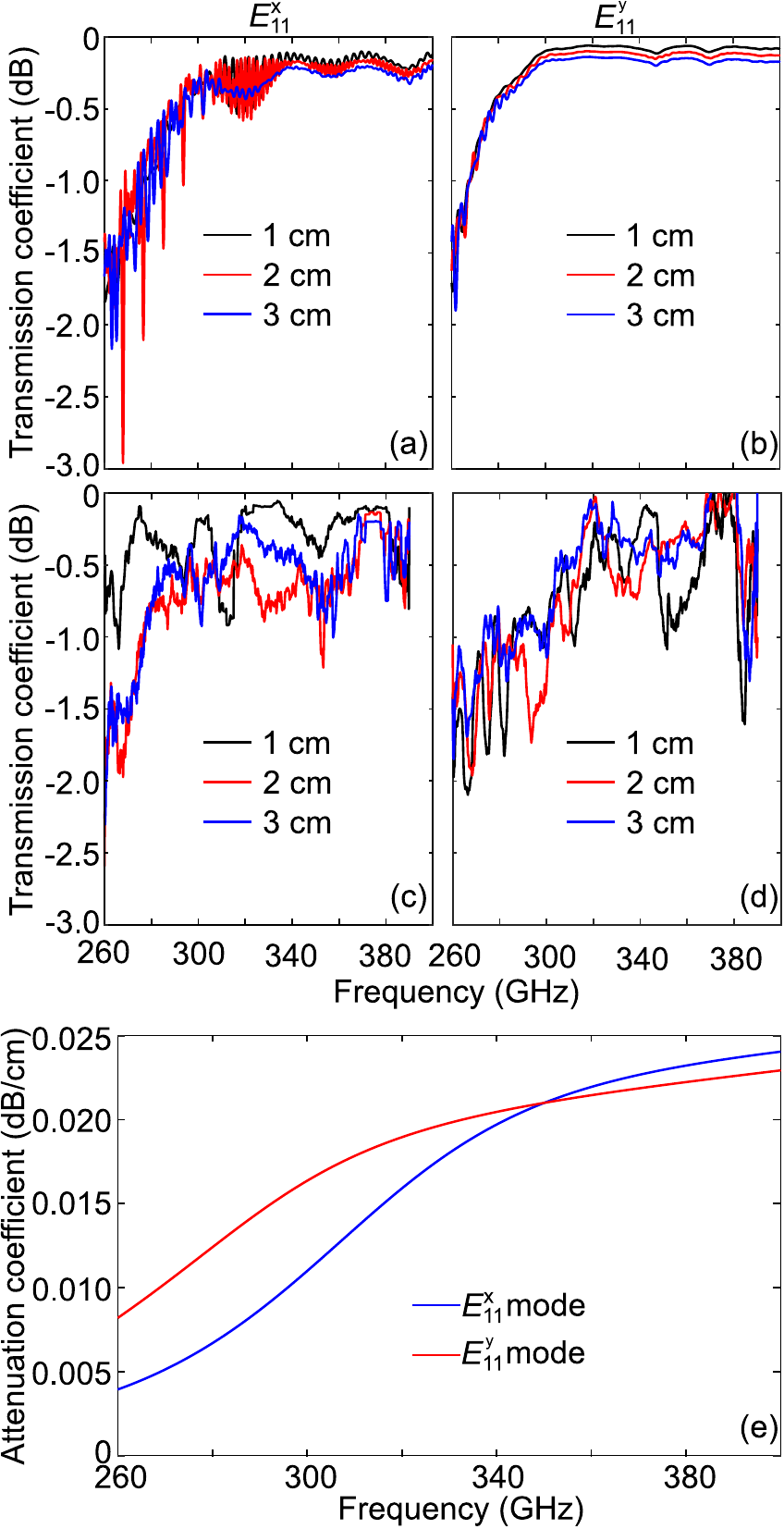}
	\caption{\textbf{Transmission and attenuation coefficients of the waveguides.} (a, b) Simulated, (c, d) measured transmission coefficients, and (e) simulated attenuation coefficient. The $E_{11}^{x}$ and $E_{11}^{y}$ modes are represented in (a, c) and (b, d), respectively. It is noted that the full-wave simulation for these waveguide lengths results in a computationally prohibitive number of mesh cells. So we approximate the cladding by a bulk material with effective permittivity tensor of $\epsilon_{x}$ = 2.75 and $\epsilon_{y}$ = 3.84. The attenuation coefficients are obtained from full-wave simulation by CST Microwave Studio with a realistic loss tangent of $3\times10^{-5}$ for the silicon.}
	\label{fig:S21}
\end{figure} 


\section*{Results}
The transmission coefficients $|S_{21}|$ of the fabricated waveguides displayed in~Fig. \ref{fig:fabrication} for both polarizations are measured with the waveguide length varying from 1 to 3 cm as shown in Fig.~\ref{fig:S21}. The measurment setups are discussed in the method section. The measurements agree well with the simulations in terms of the transmission levels. From the results shown in Fig.~\ref{fig:S21}(c) and (d), the measured transmission for the $E_{11}^{x}$ mode for each length varies from $-$2.5 to $-$0.1 dB over 260 to 390 GHz and from $-$2 to $-$0.05 dB for the $E_{11}^{y}$ mode over the band. Both the simulated and measured transmission levels are lower at the lower frequencies because of the higher coupling losses caused by the slight impedance and mode mismatches between the sample and the feed. 
Due to the evanescent fields of the dielectric waveguide, the transmission measurements are highly sensitive to the alignment with respect to the hollow waveguides. Despite the aid of micromechanical positioners in the alignment, the measured transmission profiles exhibit larger fluctuations than the waveguide attenuation. In view of this fact, we conclude that the attenuation of the waveguide is too small to be extracted accurately. However, owing to the general agreement between the simulated and measured results, it is reasonable to infer the propagation loss from the simulations. As shown in Fig.~\ref{fig:S21}(e), the attenuation coefficients excluding coupling losses are obtained from the full-wave simulation in CST Microwave Studio, where the realistic loss tangent of silicon of $3\times10^{-5}$ is adopted. The attenuation coefficient varies from 0.003 to 0.024 dB/cm for the $E_{11}^{x}$ mode, while it ranges from 0.008 to 0.023 dB/cm for the $E_{11}^{y}$ mode over 260 to 400 GHz. The loss for the $E_{11}^{y}$ mode is slightly higher over 260 to 345 GHz mainly due to its larger group  index as shown in Fig.~\ref{fig:effr}(b). 

This level of loss for the proposed effective-medium-cladded waveguides is comparable to the extremely low-loss photonic crystal waveguides~\cite{Fujita2015b,Yu2019} and bare high-resistivity dielectric waveguides~\cite{Ali2014}, which exhibited less than 0.1~dB/cm attenuation over their operation band. However,~the~loss~of~the~proposed~waveguides~is~more~than~one~order~of~magnitude~lower~than~a~commercial~WR-2.8~rectangular~waveguide ($0.287-0.436$~dB/cm).~\cite{VDI} In addition, the radiation loss caused by a $90^\circ$ bending with a radius of 2 mm is simulated in CST Microwave Studio. An average bending loss of less than 0.1~dB per bending for the two modes is achieved over 310--400~GHz and it is comparable to that of the photonic crystal waveguide~\cite{Fujita2015b} with a bending loss of about 0.2~\mbox{dB} per bending around 323--331~GHz. However, at the lower frequencies from 260 to 310 GHz, the bending loss varies from 5.5 to 0.5~dB per bending for the $E_{11}^{x}$ mode and from 3.1 to 0.1~dB per bending for the $E_{11}^{y}$ mode. The higher bending losses can be compensated by increasing the bending radius at the expense of structural compactness. 
\begin{figure}[!h]
	\centering
	\includegraphics{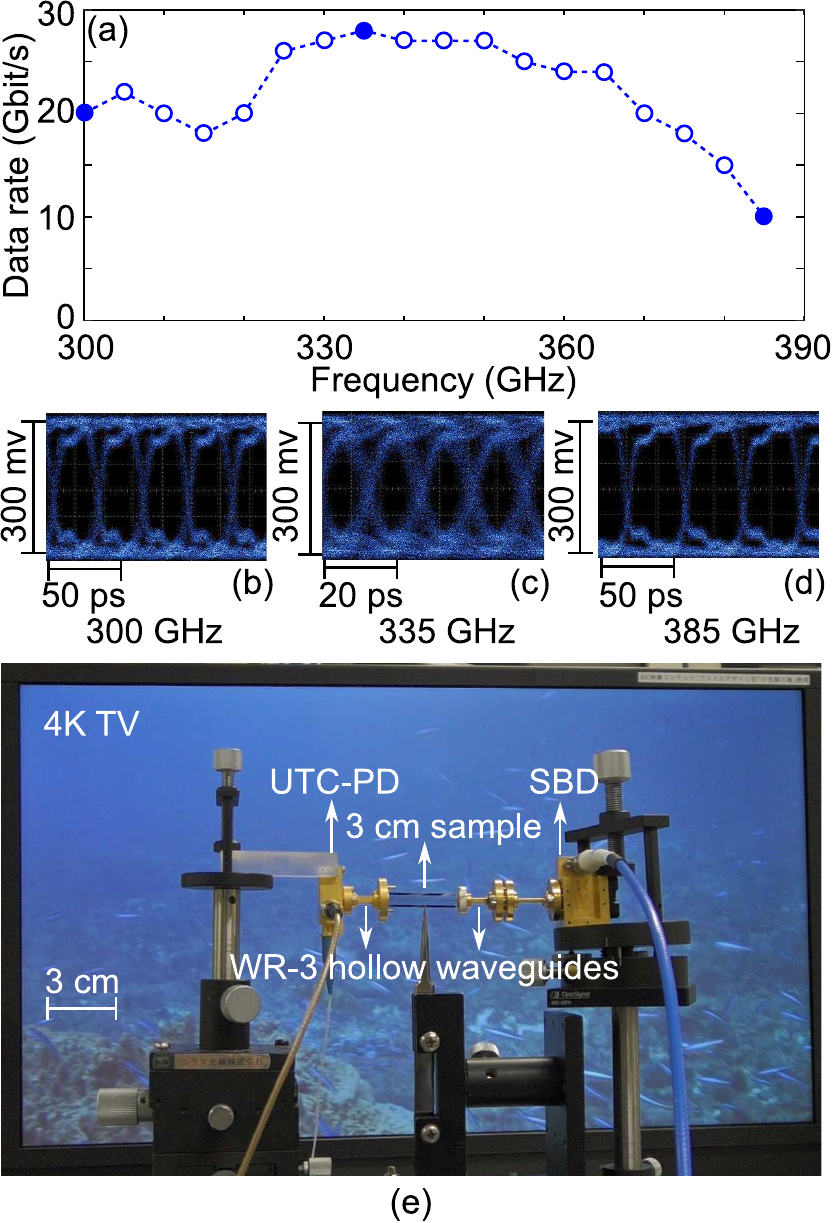}
	\caption{\textbf{Communications measurement results.} (a) Measured error-free data rate as a function of frequency over 300 to 385 GHz. The results are discrete and the lines are for visual guidance. Eye diagrams at the carrier frequencies of (b) 300 GHz (20 Gbit/s), (c) 335 GHz (28 Gbit/s), and (d) 385 GHz (10 Gbit/s). (e) Uncompressed 4K-resolution video transmission demonstration.}
	\label{fig:commu}
\end{figure}

A bit-error-rate test is conducted with a 3-cm waveguide sample for the $E_{11}^{x}$ mode over 300 to 385 GHz.~As shown in Fig.~\ref{fig:commu}(a), the error-free data rate is reasonably stable over a relatively wide bandwidth (310--370~GHz), which is attributed to the extremely low loss and low dispersion of the waveguide. At 335~GHz, the maximum measured data rate of 28 Gbit/s is achieved corresponding to the highest 3-dB bandwidth shown in Fig.~\ref{fig:effr}(c). The measurable maximum data rates are limited by the bandwidths of various components in the measurement setup. Specifically, the degradations of the data rate at lower and higher frequencies are due to the roll-off effects of the uni-travelling carrier photodiode (UTC-PD) and Schottky barrier diode (SBD). Meanwhile, the eye-diagrams have clear opening as shown in Fig.~\ref{fig:commu}(b--d) for 330, 335 and 385~GHz. An uncompressed 4K-resolution video transmission is performed as shown in Fig.~\ref{fig:commu}(e).~It~demonstrates~that~the~proposed~waveguide~can~support~a~real-time~uncompressed~4K~video~data~transmission~at~the~speed~of~6~Gbit/s. A demonstration video is available online as supplementary information.   
\section*{Conclusion and outlook}
A design of low-loss, low dispersion dielectric waveguides based on effective medium claddings has been proposed.~The waveguides are created from a single silicon wafer and do not include other materials.~The effective medium claddings enable a self-supporting and integrated \mbox{platform}. The designed waveguides can support single dominant $E_{11}^{x}$ and $E_{11}^{y}$ modes over 260 to 400~GHz, which is equivalent to a relative bandwidth of 42.4\%. The simulated attenuation coefficient varies from 0.003--0.024~dB/cm for the $E_{11}^{x}$ mode, and from 0.008--0.023~dB/cm for the $E_{11}^{y}$ mode over the operation frequency range. To characterize the transmission and communications performance, waveguide samples with various lengths have been fabricated.~The measurement shows a system-limited maximum error-free data rate of 28~Gbit/s at 335 GHz.~With demonstrated performance and form-factor, the proposed waveguide concept can be used as a platform to accommodate various passive/active devices in future terahertz wireless systems.~\cite{Withayachumnankul2018a,Yu2019a} The concept of the waveguide is applicable to operation at infrared and optical frequencies.

\section*{methods}

To validate the design, waveguide samples with lengths of 1, 2 and 3 cm have been \mbox{fabricated}. The samples are obtained from a 4-inch intrinsic float-zone silicon wafer with a thickness of 200~$\mu$m and a resistivity of 20~k$\Omega$-cm. The fabrication is based on a standard deep reactive ion-etching (DRIE) process. The waveguide core with height of 200~$\mu$m and width of 160~$\mu$m is surrounded by the effective medium claddings that are made of cylindrical air holes in a hexagonal lattice with period of 100~$\mu$m and hole diameter of 90~$\mu$m. As mentioned earlier, the solid outer claddings are for handling purposes and do not interfere with the guided waves. 
\begin{figure}[!htb]
	\centering
	\includegraphics{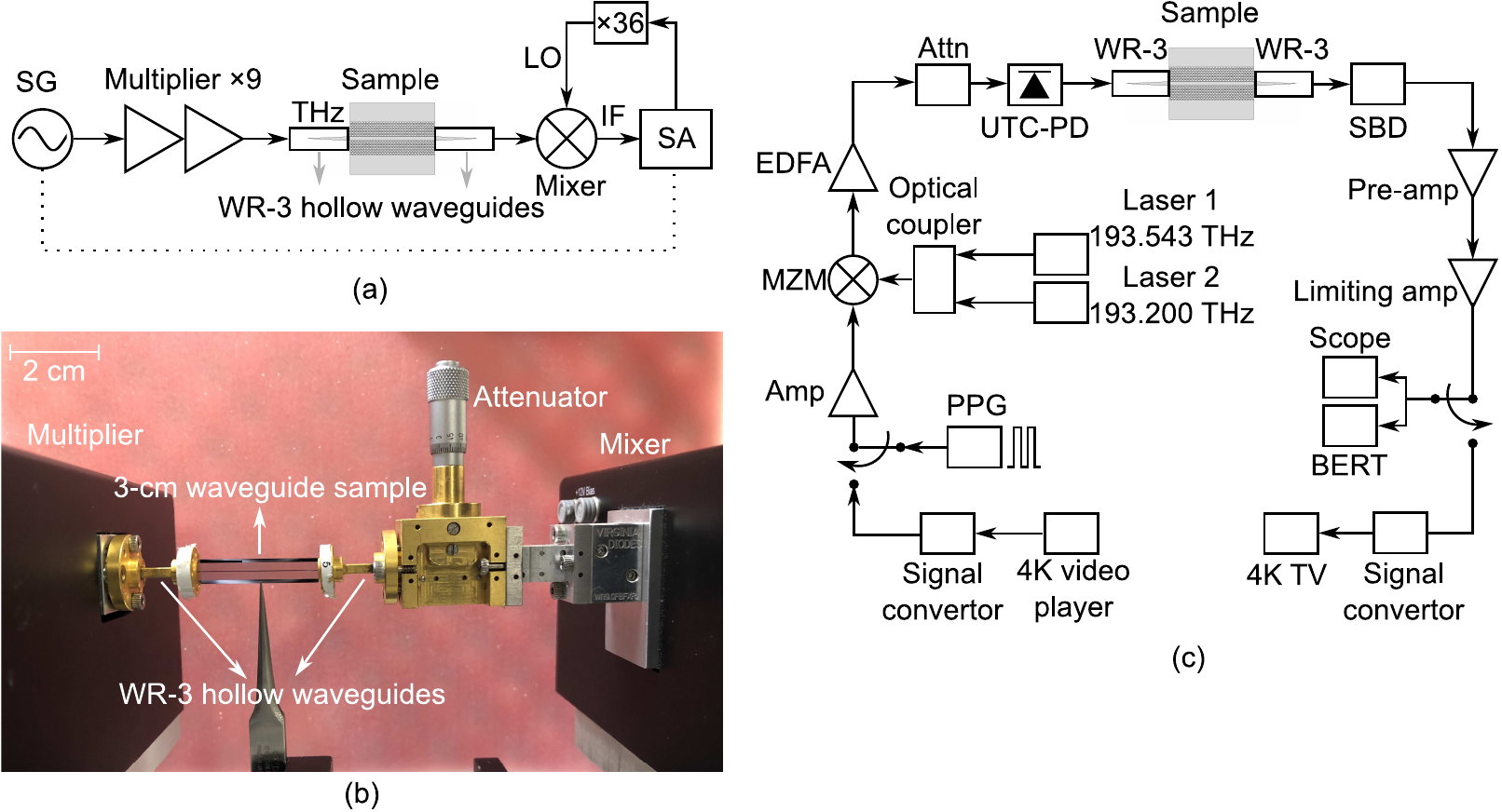}
	\caption{\textbf{Measurement setups.} (a) Transmission meassurement block diagram, (b) actual setup, and (c) communication measurement diagram. SG: signal generator, SA: signal analyser, THz: terahertz frequency, LO: local oscillator, IF: intermediate frequency, PPG: pulse-pattern generator, Amp: amplifier, MZM: Mach-Zehnder modulator, EDFA: erbium-doped fibre amplifier, Attn: attenuator, UTC-PD: uni-travelling carrier photodiode, WR-3: WR-3 hollow waveguide, SBD: Schottky barrier diode, Scope: oscilloscope, and BERT: bit-error-rate tester.}
	\label{fig:S21mea_setup}
\end{figure}
As shown in Fig.~\ref{fig:S21mea_setup}(a--b), the transmission measurements are carried out using a terahertz electronic system. The continuous wave electronic source is constructed from a millimetre-wave signal generator and a nine-fold frequency multiplier. On the receiver side, a frequency mixer downconverts the terahertz signal to the intermediate frequency (IF) at 404.4 MHz by mixing with a local oscillator (LO) signal generated from the spectrum analyser. This setup works in the frequency range from 260 to 390 GHz. The system is equipped with WR-3 hollow metallic waveguide ports.~Thus, to enable efficient transition, the tapered structures are inserted into the rectangular waveguides, so that the modes can couple between the hollow waveguides and the sample. In this way, a dielectric waveguide sample positioned between the transmitter and the receiver can be measured with minimal coupling losses. 

To validate the waveguide performance in terahertz communications, bit-error-rate testing experiments are performed and an uncompressed 4K-resolution video transmission is demonstrated. A diagram of the measurement system is shown in Fig~\ref{fig:S21mea_setup}(c). On the transmitter side, two tunable near-infrared laser sources generate a beat optical signal that is modulated with the signal from a pulse-pattern generator or a 4K-resolution video player. Amplified by the EDFA, the modulated signal is down-converted to the terahertz signal by a UTC-PD and then coupled to the dielectric waveguide through a WR-3 hollow waveguide. On the receiver side, a SBD is used to retrieve the modulating data from the received terahertz signal. The data signal is then reshaped by a preamplifier and a limiting amplifier. The data quality is then measured by a bit-error rate tester and an oscilloscope, while for the uncompressed 4K-resolution video transmission measurement, the signal can be displayed on a 4K television after a digital signal converter.

\begin{acknowledgments}
	We wish to acknowledge support from the Australian Research Council Discovery Project (No. ARC DP180103561), and the Core Research for Evolutional Science \& Technology (CREST) program of the Japan Science and Technology Agency (JST) (\#JPMJCR1534). The Quadro P6000 GPU used for this research was donated by NVIDIA Corporation to the Terahertz Engineering Laboratory, The University of Adelaide.
\end{acknowledgments}

\section*{Author Contributions}
W. G. designed the waveguides and wrote the manuscript with support from C. F., W. W., and M. F. in the analysis. X. Y. performed experiments. T. N. designed and supported the communications experiments. W. W. conceived and coordinated the project with M. F. All authors contributed to the manuscript.

\section*{Competing Interests Statement}
The authors declare no competing financial interests. Readers are welcome to comment on the online version of the manuscript. Correspondence and requests for materials should be addressed to Withawat~Withayachumnankul (withawat@adelaide.edu.au) and Masayuki Fujita (fujita@ee.es.osaka-u.ac.jp).

\bibliographystyle{nature}
\bibliography{WG1}
 
\end{document}